# Analyzing β-Lactamase Evolution from the Principle of Least Action Perspective

Pablo Garay[1], Walter J. Lapadula[2], Maximiliano Juri Ayub[2], and Jorge A. Vila[1]

[1] Institute of Applied Mathematics San Luis (IMASL), National University of San Luis (UNSL), and National Scientific and Technical Research Council (CONICET), Ejército de Los Andes 950, 5700 San Luis, Argentina.
[2] Multidisciplinary Institute of Biological Research (IMIBIO), National University of San Luis (UNSL), and National Scientific and Technical Research Council (CONICET), Ejército de Los Andes 950, 5700 San Luis, Argentina.

Protein sequences change over time due to the accumulation of mutations in the genes that encode them. Nonetheless, accurately identifying the most probable evolutionary pathways—and more efficient trajectories—between an ancestral and a derived sequence remains a challenge. This difficulty stems from a limited understanding of the critical factors that drive the evolutionary process. This study aims to address this issue by validating a newly proposed mechanistic model—grounded on the principle of least action—for analyzing protein evolution, using β-lactamase as an empirically characterized evolutionary model. The initial findings indicate that, after accounting for how mutations—and the context-dependent interactions between them (epistasis)—affect protein stability and, hence, the kinetics of protein folding, a resort to the principle of least action allows us to simultaneously identify the most probable pathways and the most efficient trajectories swiftly and accurately. These findings also suggest that the most probable evolutionary protein pathways are the ones that exhibit the most efficient trajectories. All in all, our study addresses several unanswered questions regarding the main factors that govern how protein evolves at the molecular level—under constant selection pressure—and outlines directions for future research in key fields such as directed evolution and ancestral sequence reconstruction.

**Introduction**

The issue of protein sequence evolution through single-point mutations in the encoding gene—resulting in amino acid replacements that connect an ancestral sequence (wild type) to a derived sequence through a variable number of mutational steps—was rationalized by John Maynard Smith in his seminal 1970 article: "…*If evolution by natural selection is to occur, functional proteins must form a continuous network that can be traversed by unit mutational steps without passing through nonfunctional intermediates*…" This proposal implies that any functional protein within the protein sequence space must conform to the thermodynamic hypothesis or Anfinsen's dogma (Anfinsen, 1973); consequently, prions and intrinsically disordered proteins will be omitted from this analysis. Additionally, in this evolutionary model, if the ancestral and derived protein sequence remain unchanged throughout the process, it operates akin to a word game (Maynard Smith, 1970). In this game, a word is transformed into another by substituting one letter at a time; for example, altering the word 'NCSI' to 'IRNL' (Ogbunugafor, 2020), where each letter represents an amino acid in the single-letter code. Thus, each sequence of amino acid substitutions defines an evolutionary pathway and, consequently, a trajectory. The identification of the most probable pathways and efficient trajectories among all feasible alternatives is one of the primary challenges that this evolutionary model faces. Under the simplifying assumption that all mutational differences between the ancestral and derived sequences are known and that each mutation occurs only once, the number of mutational steps is the primary factor determining the number of possible pathways. For instance, if there are $n$ mutational steps, there will be $n!$ possible pathways connecting the wild type to the derived sequence. More specifically, evaluating the 120 pathways that link two protein variants separated by five-point mutations is crucial for determining the most probable pathway of a true evolutionary process such as that of β-lactamase, which increases its activity under selective pressure when exposed to cefotaxime (Weinreich *et al*., 2006). It is worth noting that Weinreich *et al*.'s (2006) research—which experimentally constructed a fitness landscape connecting those β-lactamase variants—has stimulated investigations into the mechanistic modeling (Glennan, 2017, Craver, 2025) of adaptive evolution and continues to engage with applications in biomedicine, biological engineering, and artificial life (Ogbunugafor, 2026). Besides all the foregoing considerations, the rationale for selecting β-lactamase as an empirically characterized evolutionary model system—to validate the results obtained using a recently introduced evolutionary model (Vila, 2026)—is grounded in several other important observations. Among all of them, we select the following: (*i*) the most common mechanism of resistance to β-lactam antibiotics in Gram-negative bacteria is the production of β-lactamases that hydrolyze the drugs; (*ii*) β-lactam antibiotics are the most often-used antimicrobials, representing ~65% of antibiotic usage worldwide (Livermore & Woodford, 2006); (*iii*) β-lactamases are thought to have evolved from penicillin-binding proteins and are ubiquitously present in the chromosomes of most Gram-negative bacteria (Fröhlich *et al*., 2021); (*iv*) The X-ray structure of *Escherichia coli* TEM1 β-lactamase—a protein of 263 amino acids—has been refined (PDB ID 1BLT) at 1.8 Å resolution (Jelsch *et al*., 1993); (*v*) It has been shown that five-point mutations—in a particular β-lactamase allele—jointly increase bacterial resistance to cefotaxime by a factor of ~$10^5$ and that, among other important results, 102 out of 120 mutational trajectories are selectively inaccessible (Weinreich *et al*., 2006). The aforementioned information on β-lactamase, along with a recently proposed methodology grounded in the principle of least action (Vila, 2026), will be employed to validate the identification of the most probable evolutionary trajectories and the primary factors influencing this process, as well as to compare the obtained results with those of Weinreich *et al*. (2006). At this point, it is important to clarify two key points before we proceed.

Firstly, the suggested comparison does not seek to question the analysis of Weinreich et al. (2006). Instead, it attempts to assess the merits and drawbacks of the newly proposed methodology—grounded in the principle of least action—for analyzing protein evolutionary pathways and trajectories, investigating the origins of potential discrepancies arising from the comparison, identifying its limitations, and suggesting possible solutions to address these issues. Secondly, the objective of introducing a mechanistic model (Glennan, 2017, Craver, 2025) is to outline a sequential process that clarifies how protein evolution occurs under constant selection pressure. This model examines how mutations affect protein stability within the framework of protein evolution. Its goal is to identify the key biophysical factors that might influence the fitness effects of these mutations and, in turn, shape evolutionary pathways and trajectories. While protein stability is regarded as a primary biophysical determinant of protein fitness, it is acknowledged that other factors may also play a role. It is important to note that our analysis does not consider other evolutionary forces—such as genetic drift, changing selective pressures, demographic effects, horizontal gene transfer, recombination, or historical contingencies. This omission is intentional, as including these factors would exceed the scope of this paper.

To achieve our objectives, we will start with a comprehensive description of the proposed methodology for identifying the most probable evolutionary pathways and trajectories of proteins in general, with a specific focus on β-lactamases. This will be supplemented by a clear outline of the fundamental assumptions and approximations that underlie the analysis.

**Overview of the β-lactamase evolutionary model used for testing**

Weinreich *et al*. (2006) studied the evolution of increased cefotaxime resistance by thoroughly examining, both experimentally and theoretically, the effects of five specific mutations.

However, the analysis presented here is purely theoretical and focuses on four missense mutations identified by Weinreich *et al*. (2006): A42G, E104K, M182T, and G238S. In this notation, the letters denote amino acids using the one-letter code, whereas the numbers indicate their positions within the sequence of *Escherichia coli* TEM-1 β-lactamase (PDB ID 1BLT). We omit the fifth mutation reported by Weinreich *et al*. (2006), g4205a, because it is located outside the coding region and therefore cannot be evaluated using the protein stability–based framework employed in this study.

All in all, the theoretical analysis we conducted on the development of resistance to β-lactam antibiotics—which focuses on four missense mutations—does not undermine or limit our ability to compare our findings with those of Weinreich *et al*. (2006) following the same number and type of mutations.

**A mechanistic model to examine how mutations affect the evolution of proteins**

Before we begin our mathematical description of a mechanistic model aimed at analyzing β-lactamase evolution, it is important to clarify two key points. Firstly, the distinction between pathways and trajectories is crucial. Although this distinction has been defined previously (Vila, 2026), we will reiterate it here for the reader's convenience, as the terms "pathway' and "trajectory" represent distinct concepts. A pathway is a spatial structure made up of a series of steps, each of which represents a single-point mutation. However, it does not provide any information about time. In contrast, a trajectory is a pathway that includes temporal data, specifying the speed at each step and the duration of movement along the path. From this perspective, the term "efficiency," which refers to optimal output in relation to time, is specifically applicable to the classification of trajectories. Conversely, the term "probable" is utilized for classifying pathways, as time is not a relevant factor in this context. Secondly, the mechanistic model used to analyze β-lactamase evolution—discussed below—is grounded in the principle of least action (Vila, 2026). Nevertheless, this physical principle is employed as a theoretical assumption—rather than as a strict physical-mathematical description of the proteins' evolutionary problem—that enabled us to determine the most likely pathways and the most efficient trajectories between the ancestral and derived sequences. The reasons to adopt such an approach are multiple, the main one being that a strict application of the least action principle would lead to a complex physical-mathematical problem (Coopersmith, 2017) whose solution is far from trivial, as is the case of the protein folding problem (Vila, 2026b).

We have recently proposed (Vila, 2026) that in the presence of an arbitrary number of possible evolutionary pathways ($\mu$)—connecting an ancestral protein with a derived sequence after a fixed number of mutations ($j$)—the most probable trajectory is simply the most efficient, namely, the one that minimizes the 'action' $\Gamma_j^{\mu}$ (a functional of the pathway) defined as follows:

$$\Gamma_{\mathrm{j}}^{\mu} = \sum_{k=1}^{\mathrm{j}} \tau_k^{\mu} = \tau_{\mathrm{wt}} \sum_{\mathrm{k}=1}^{\mathrm{j}} \left(\tau_{\mathrm{k}}^{\mu}/\tau_{\mathrm{wt}}\right) = \tau_{\mathrm{wt}} \sum_{k=1}^{\mathrm{j}} e^{\beta\Delta\Delta G_k^{\mu}} \approx \tau_{\mathrm{wt}} \sum_{k=1}^{\mathrm{j}} e^{\beta\Delta\Delta G_{k,U}^{\mu}} \qquad \text{(sec) (1)}$$

where $\tau_k^{\mu} \sim \tau_0\, e^{\beta\Delta G_k^{\mu}}$ indicates the unfolding/folding rate, *i.e.*, the time required to overcome the free-energy barrier $\Delta G_k^{\mu}$ (Vila, 2023), which reflects the protein's marginal stability upon the $k$-mutation (Vila, 2022). This free-energy barrier is characterized by a set of high-energy native-like structures, known as the transition state ensemble (Akmal & Muñoz, 2004). These structures coexist in rapid dynamic equilibrium with the native state (Hormoz, 2013; Vila, 2023), representing a threshold beyond which the protein unfolds or becomes nonfunctional (Martin & Vila, 2020); $\tau_0$ is a pre-exponential factor that represents the unfolding/folding speed limit of two-state proteins, which is also referred to as the barrier-less limit (Eaton, 2021); $\tau_{wt} \sim \tau_0\, e^{\beta \Delta G_{wt}}$ indicates the time required to overcome the wild-type free-energy barrier $\Delta G_{wt}$; $j$ designates the final mutational step of the evolutionary trajectory, and therefore $j!$ represents the total number of possible pathways ($\mu$); $\beta$ is defined as $1/RT$, where $R$ stands for the gas constant, and $T$ represents the absolute temperature in Kelvin. $\Delta\Delta G_k^{\mu} = (\Delta G_k^{\mu} - \Delta G_{wt})$ represents the 'protein marginal stability change' between the wild-type (*wt*) and the derived sequence (*k*), while $\Delta\Delta G_{k,U}^{\mu} = (\Delta G_{k,U}^{\mu} - \Delta G_{wt,U})$ represents the 'protein unfolding Gibbs free-energy change' (Bigman & Levy, 2018) between the wild-type (*wt*) and the derived sequence (*k*). In Eq. (1), the approach $\Delta\Delta G_k^{\mu} \approx \Delta\Delta G_{k,U}^{\mu}$—where the latter can be determined by high-throughput methods, such as cDNA display proteolysis (Tsuboyama *et al*., 2023)—represents a suitable strategy for reliably assessing the impact of point mutations on protein marginal stability change (Vila, 2022; Vila, 2026). Consequently, the observed change in the unfolding Gibbs free energy ($\Delta\Delta G_{k,U}^{\mu}$) should fundamentally reflect the change in the protein marginal stability ($\Delta\Delta G_k^{\mu}$). Therefore, for a given evolutionary problem, such as for β-lactamases, the starting point of any evolutionary trajectory $\mu$ is that of the wild-type; then the problem to determine the most probable trajectory will be reduced to computing all possible values for the redefined dimensionless functional $\Gamma_{\mu,j}^{*}$ $(\frac{\Gamma_j^{\mu}}{\tau_{\mathrm{wt}}})$, which is defined as:

$$\Gamma_{\mu,\mathrm{j}}^{*} \approx \sum_{k=1}^{\mathrm{j}} e^{\beta \Delta\Delta G_{k,U}^{\mu}} \tag{2}$$

This equation applies to evolutionary pathways, characterized by obeying a Markovian process—a time-independent stochastic process in which the probability of each step depends solely on the preceding step—arising from the nature of the epistatic interactions (Vila, 2024) that govern the order in which mutations occur (Vila, 2026).

From the perspective of the least action principle (Vila, 2026), the most efficient trajectory is the one for which the functional ($\Gamma^{*}_{\mu,j}$) satisfies the following condition: $\delta\Gamma^{*}_{\mu,j} = 0$. In other words, the most probable trajectory must be the one for which an infinitesimal variation (δ) on the protein marginal stability—such as those caused by changes in the order in which the mutations occur—corresponds to a stationary point of the functional, which must be at a minimum or a saddle point but not at a maximum. As evident from Eq. (2), destabilizing mutations ($\Delta\Delta G^{\mu}_{k,U} < 0$) reduce the contribution to Eq. (2), thereby minimizing the value of the functional or 'action.' Consequently, the most evolutionarily efficient trajectory—regardless the number of potential mutations—should incorporate as many destabilizing mutations as feasible. A cautionary note is warranted at this juncture. Throughout the selective evolutionary process, any mutation—except for reasons evident from Eq. (2), such as non-coding mutations like g4205a in β-lactamase—is permissible as long as the following condition is met: $|\Delta\Delta G| < \sim 7.4$ kcal/mol (Martin & Vila, 2020). This threshold is determined by the validity of the Anfinsen dogma (Vila, 2022); if it is exceeded, the protein may unfold or lose its functionality (Vila, 2026). Therefore—under this constraint—the trajectory that minimizes the function delineated in Eq. (2) is expected to be the most evolutionary efficient. This trajectory generally comprises a blend of stabilizing and destabilizing mutations, as will be shown subsequently for β-lactamase.

Before moving forward, it's essential to clarify two key points. Firstly, each sequence of mutations creates a unique pathway, and the calculation in Eq. (2) yields the value of the trajectory associated with that specific pathway. Therefore, we must evaluate the possibility that multiple pathways may lead—particularly if there is a reasonable margin of uncertainty in the

calculations—to trajectories similar to the most probable one. If this scenario arises, one could argue that, rather than a single most probable pathway, we should consider a set of pathways with similar probabilities. The latter may not indicate a limitation of the method; rather, it could emphasize an evolutionary advantage of the system under study, as multiple pathways might provide effective solutions to the evolutionary challenge. Our results, which we will address in the Results and Discussion section, indicate that this possibility is indeed viable. Secondly, calculating Eq. (2) allows us to identify the most efficient trajectories and the most probable evolutionary pathways simultaneously.

***Methodology***

To identify the most probable trajectory among the 24 possible pathways from $TEM^{wt} \rightarrow$ TEM* for β-lactamase (Weinreich *et al*., 2006), which arise from four missense mutations (A42G, E104K, M182T, and G238S), we must evaluate Eq. (2) for each single-point mutation in the specified order (see Table 1). For this purpose, the calculation of all possible $\Delta\Delta G_{k,U}^{\mu}$ was carried out by using the FOLDX force field (Delgado *et al*., 2025) with the PDB structure 1BLT for β-lactamase wild-type (*wt*) structure as input. This software was chosen because it is recognized as one of the leading standards for predicting ΔΔG in proteins (Delgado *et al*., 2025). FoldX utilizes a rigid-body approach, meaning the backbone remains unchanged during mutations. Consequently, the ΔΔG free energy calculations for each mutational step across the 24 pathways (shown in Table 1) are not a single calculation; instead, they are computed as average values (<ΔΔG>) from three separate runs of minimizing the total free energy of the force field. This feature of FoldX was designed to improve the efficiency of conformational sampling after a mutation (Delgado *et al*., 2025). The choice to conduct three runs, for each mutation, represents a balance between accuracy and calculation speed. This averaging technique helps alleviate any potential side-chain atomic overlaps that may arise from each single-point mutation. All calculations of Eq. (2)—shown in Table 1—were performed at room temperature (T = 298 K).

To reproduce the results listed in Table 1, it is important to clarify a key point. The derivation of Eq. (2) is based on the convention that $\Delta\Delta G = (\Delta G_{m} - \Delta G_{wt}) < 0$ indicates a destabilizing mutation (Vila, 2023; Vila, 2026). In contrast, the computation of ΔΔG using the force field FoldX (Delgado *et al*., 2025) assumes that $\Delta\Delta G < 0$ indicates a stabilizing mutation. At this point, we face two options: we can either adopt the FoldX convention and change the sign of the argument in the exponential of Eq. (2) to align with the principle of least action, or we can maintain our original convention and leave Eq. (2) as it was initially formulated. To maintain consistency with prior analyses, we have opted for the latter option; consequently, Table 1 adheres to this convention. The calculations were performed using an Intel Core i7 CPU with 24 GB of RAM. The total time required to complete all 24 trajectories listed in Table 1 was 97 minutes.

**Results and Discussion**

The analysis depicted in Figure 1 shows a ranking of the most evolutionary efficient trajectories among the 24 that result from four missense mutations in β-lactamase. The trajectory identified as the most probable—the most efficient one, according to our calculations—consists of the following sequence of mutations: G238S, A42G, M182T, and E104K. The trajectory associated with this evolutionary pathway—grounded in the functional value obtained from Eq. (2)—align closely with the predictions of Weinreich *et al*. (2006) for the most probable pathway, namely, the one defined by the mutation sequence: G238S, E104K, A42G, and M182T. The inset of Figure 1 illustrates the six pathways identified as the most probable, as calculated using Eq. (2). A noteworthy result—among these six ones—is the one obtained for the pathway with the following sequence of mutations: G238S, A42G, E104K, and M182T. This evolutionary pathway represents the second most probable one according to both Weinreich *et al*. (2006) and our calculations, using Eq. (2). This point is highlighted in green in Figures 1 and 2. In particular, Figure 2 illustrates the variations in free energy resulting from single-point mutations across the 24 evolutionary pathways. This figure particularly highlights the differences among three likely pathways: two identified by Weinreich *et al*. (2006) and the one deemed most probable based on our calculations using Eq. (2). The minor discrepancies among the six most efficient trajectories, as illustrated in Table 1 and Figure 1, contrast sharply with the values of the other 18 trajectories. This outcome prompts us to conclude, as previously mentioned, that multiple equally efficient evolutionary trajectories may be present. This suggests that numerous pathways, rather than a single one, may efficiently direct the evolution of β-lactamases towards the derived sequence.

Finally, we must acknowledge a compelling observation regarding a significant mutation that characterizes the six most probable pathways previously discussed. Specifically, we refer to G238S, which denotes the initiation of each of these pathways (see Figure 1). This mutation shows the highest increase in cefotaxime resistance on $TEM^{wt}$, according to Weinreich *et al*. (2006), and it exhibits the largest destabilizing effects among all possible starting mutations (see Table 1 and Figure 2). Nevertheless, not all mutations can be destabilizing, as shown in Table 1 and Figure 2. In fact, the balance between stabilizing and destabilizing mutations reflects the interplay between evolutionary efficiency and the thermodynamic constraints governing protein structural stability (Vila, 2024b). Within this framework, this balance is governed—at the molecular level—by the validity of the thermodynamic hypothesis (Anfinsen, 1973), which establishes a threshold for permissible changes in free energy resulting from mutations ($|\Delta\Delta G| \sim 7.4$ kcal/mol). Beyond this threshold, a protein may unfold or lose its functionality (Martin & Vila, 2020).

The findings from this study focus on identifying the most efficient evolutionary trajectories and, hence, the most probable evolutionary pathways of β-lactam resulting from mutations—specifically, examining how these mutations alter the protein's marginal stability. Therefore, it is essential to select appropriate computational methods to accurately measure these changes. In this context, the selection of an appropriate force field, such as FoldX, resulted in a sizable agreement with the findings of Weinreich *et al*. (2006). This outcome offers strong support for both the trajectory calculations outlined in Eq. (2) and the idea that the principle of least action may serve as a valuable framework for analyzing evolutionary trajectories within this system.

Overall, our findings are consistent with those of Weinreich *et al*. (2006) regarding the most probable set of pathways in β-lactamase evolution and with the most efficient trajectories—associated with fast evolution—of Ogbunugafor (2026), which is encouraging. However, the impact of noncoding mutations and other previously overlooked factors, such as alterations in the surrounding environment—such as pH, ionic strength, temperature, *etc*.—underscores certain limitations in the simplification of evolutionary process modeling. Nonetheless, tackling these factors is beyond the purview of this study. Therefore, the proposed method could be employed to predict the most efficient evolutionary trajectories and to assess the relative contributions of those factors—as well as various biophysical constraints or evolutionary mechanisms outside the scope of this study. The relative importance of these contributions is expected to vary substantially among proteins, surrounding environments, and evolutionary scenarios.

**Conclusions**

The consistency between the predictions derived from the principle of least action and the experimental results reported for β-lactamase evolution suggests that our approach may provide a useful mechanistic model for analyzing protein evolutionary pathways and trajectories. This model outlines a sequential process that clarifies how protein evolution under selection pressure occurs by analyzing the effects of mutations on protein stability. In particular, it highlights the key biophysical factors that influence the fitness effects of mutations. For instance, the impact of the sequence of mutations emphasizes the Markovian nature of the evolutionary process, which, together with epistatic effects (Vila, 2024; 2026), shapes both the most evolutionarily efficient trajectories and the most probable pathways. Concerning the identification of the latter, it is important to highlight that while Weinreich *et al*. (2006) concentrated on experimentally and theoretically enumerating mutational pathways and quantifying their selective accessibility, Ogbunugafor (2026) focused on ascertaining whether the most probable pathways represent the fastest trajectories. In contrast, our mechanistic model for analyzing protein evolution enabled us to address both issues simultaneously. Indeed, the solution of Eq. (2) enables us to identify the most efficient evolutionary trajectories and the most probable pathways concurrently. The alignment of our findings with the observation of Weinreich *et al*. (2006) and Ogbunugafor (2026) serves as a strong validation of the physical foundation on which our approach is grounded. At this stage, we might ask, what is the significance of a reliable mechanistic description of the proteins' evolutionary process? A compelling answer is offered by Liberles *et al*. (2012), who stated, "…*The evolution of biomacromolecules is complex, and there is a constant tension between generating simple models and embracing the complexity of molecular evolution. As models that describe mechanistic processes and fit data well/offer explanatory power are generated, our corresponding* *understanding of protein evolution and protein biophysics will increase...*" Based on these reflections, we conclude that our analysis of protein evolution has enhanced our understanding of this complex topic. For instance, our research has revealed that changes in protein marginal stability—and consequently in the kinetics of the folding process—serve as a highly sensitive probe with which to simultaneously identify, based on the principle of least action validity, the most efficient trajectories and the most probable evolutionary pathways of proteins. This achievement highlights that protein stability is essential for the validity of the principle of least action. Indeed, the most efficient evolutionary trajectories require the emergence of destabilizing mutations—that speed up the protein unfolding/folding process—thereby increasing the probability of a specific evolutionary pathway. At this point, it is important to note that most mutations in proteins, namely for ~70% of their sites, are destabilizing (Vila, 2024b), thereby ensuring the necessary conditions—at a molecular level—for the appearance of efficient evolutionary processes under constant selection pressure. This conclusion is supported by evidence indicating that protein stability plays a significant role in facilitating evolution (Bloom *et al*., 2006; Tokuriki & Tawfik, 2009; Bloom & Arnold, 2009; Arnold, 2009; Romero & Arnold, 2009; Ota *et al*., 2018).

In general, our results—which are based on the validity of the principle of least action—lead us to propose which are the most probable evolutionary pathways. This conjecture, on one hand, aligns with a proposed solution to a long-standing issue in structural biology: how proteins fold. In fact, the solution implies that proteins should adopt the most efficient folding pathways rather than exploring the whole accessible conformational space (Vila, 2026b). On the other hand, it backs up the idea proposed by Weinreich *et al*. (2006), which states that "replaying the protein tape of life might be surprisingly repetitive," at least under constant selection pressure.

Last, but not least, it is important to acknowledge that the applications of the proposed methodology for predicting the most probable evolutionary pathways and trajectories of proteins may hold considerable significance in two pertinent fields of evolutionary biology. These fields specifically refer to “directed evolution” (Arnold, 2009; Bloom & Arnold, 2009; Romero & Arnold, 2009) and “ancestral sequence reconstruction” (Joy *et al*., 2016; Spence *et al*., 2021; Puente-Lelievre *et al*., 2025). In the context of directed evolution, our method could help better understand experimental outcomes by identifying the mutational trajectories that are expected to be most accessible under stability-based biophysical constraints. Conversely, directed evolution experiments offer an ideal system for testing the validity and predictive power of the present model because they generate evolutionary trajectories under controlled selective conditions. On the other hand, for ancestral sequence reconstruction, the proposed approach could be useful to prioritize among alternative ancestral sequence candidates by identifying the most likely mutational routes connecting ancestral and descendant proteins. Therefore, the method could function as a “biophysical-based filter” that helps prioritize alternative ancestral reconstructions, directing experimental gene synthesis towards the most promising candidates. Further progress on these topics will be published elsewhere.

**Acknowledgments**

The author acknowledges the National University of San Luis (UNSL), the National Scientific and Technical Research Council (CONICET), the Institute of Applied Mathematics San Luis (IMASL), and the Multidisciplinary Institute of Biological Research (IMIBIO) for their assistance. W.J.L. and M.J.A. are members of the Scientific Career of CONICET. The authors want to call the attention of the international scientific community to the profound crisis currently affecting Argentina's science and technology system as a consequence of the severe funding cuts and institutional policies implemented by the administration of President Javier Milei. The authors are deeply concerned that these measures threaten the sustainability of scientific research, higher education, and Argentina's long-term capacity for innovation.

**Code Availability**

The code for reproducing the results and the associated software and hardware requirements is available at https://github.com/BIOS-IMASL/pathways_of_evolution.

**Funding**: This work was written without any financial support.

**Conflict of interest/ Competing interests:** None.

**Ethics approval**: Not applicable

**Table 1**

| Pathway | $\langle\Delta\Delta G_1\rangle$ | $\langle\Delta\Delta G_2\rangle$ | $\langle\Delta\Delta G_3\rangle$ | $\langle\Delta\Delta G_4\rangle$ | $\Gamma^*_{\mu-j}$ |
|---|---|---|---|---|---|
| GS-AG-MT-EK | -2.51264 | -2.66453 | -3.26006 | -2.72600 | 0.03953 |
| GS-AG-EK-MT | -2.51264 | -2.66453 | -2.53119 | -2.72600 | 0.04938 |
| GS-MT-AG-EK | -2.51264 | -2.11978 | -3.16361 | -2.59725 | 0.05944 |
| GS-EK-AG-MT | -2.51264 | -1.95838 | -2.53119 | -2.72600 | 0.07487 |
| GS-MT-EK-AG | -2.51264 | -2.11978 | -2.04148 | -2.59725 | 0.08647 |
| GS-EK-MT-AG | -2.51264 | -1.95838 | -2.04148 | -2.59725 | 0.09520 |
| AG-GS-MT-EK | -0.72880 | -2.66453 | -3.26006 | -2.72600 | 0.31719 |
| AG-GS-EK-MT | -0.72880 | -2.66453 | -2.53119 | -2.72600 | 0.32704 |
| AG-MT-GS-EK | -0.72880 | -0.80206 | -3.26006 | -2.72600 | 0.56412 |
| AG-EK-GS-MT | -0.72880 | -0.47195 | -2.43987 | -2.73492 | 0.76873 |
| AG-MT-EK-GS | -0.72880 | -0.80206 | -0.59651 | -2.73492 | 0.92503 |
| MT-GS-AG-EK | -0.05828 | -2.11978 | -3.16361 | -2.59725 | 0.95134 |
| MT-GS-EK-AG | -0.05828 | -2.11978 | -2.04148 | -2.59725 | 0.97837 |
| AG-EK-MT-GS | -0.72880 | -0.47195 | -0.59651 | -2.73492 | 1.11762 |
| MT-AG-GS-EK | -0.05828 | -0.83212 | -3.16361 | -2.59725 | 1.16873 |
| MT-AG-EK-GS | -0.05828 | -0.83212 | -0.62452 | -2.77336 | 1.50900 |
| EK-GS-AG-MT | 0.25685 | -2.01809 | -2.43987 | -2.73492 | 1.60233 |
| EK-GS-MT-AG | 0.25685 | -2.01809 | -2.17635 | -2.77336 | 1.61081 |
| EK-AG-GS-MT | 0.25685 | -0.47195 | -2.43987 | -2.73492 | 2.01987 |
| MT-EK-GS-AG | -0.05828 | 0.14448 | -2.17635 | -2.77336 | 2.21719 |
| EK-AG-MT-GS | 0.25685 | -0.47195 | -0.59651 | -2.73492 | 2.36876 |
| MT-EK-AG-GS | -0.05828 | 0.14448 | -0.62452 | -2.77336 | 2.54012 |
| EK-MT-GS-AG | 0.25685 | 0.14448 | -2.17635 | -2.77336 | 2.85409 |
| EK-MT-AG-GS | 0.25685 | 0.14448 | -0.62452 | -2.77336 | 3.17702 |

This table provides comprehensive information on 24 possible evolutionary pathways linked to four missense mutations in β-lactamase. The first column lists these pathways, where, for simplicity, we have not included the specific positions of the mutations in the β-lactamase sequence. Columns two to five show the average free energy changes ($\langle\Delta\Delta G\rangle$) for each mutation, which were obtained from three FoldX calculations, as explained in the main text. The sixth column presents the computed functional values for each of the 24 corresponding trajectories, as determined by Equation (2). The values for the functional are organized in ascending order, with the minimum value (representing the most efficient trajectory) at the top and the maximum value (indicating the least efficient trajectory) at the bottom, according to Equation (2).

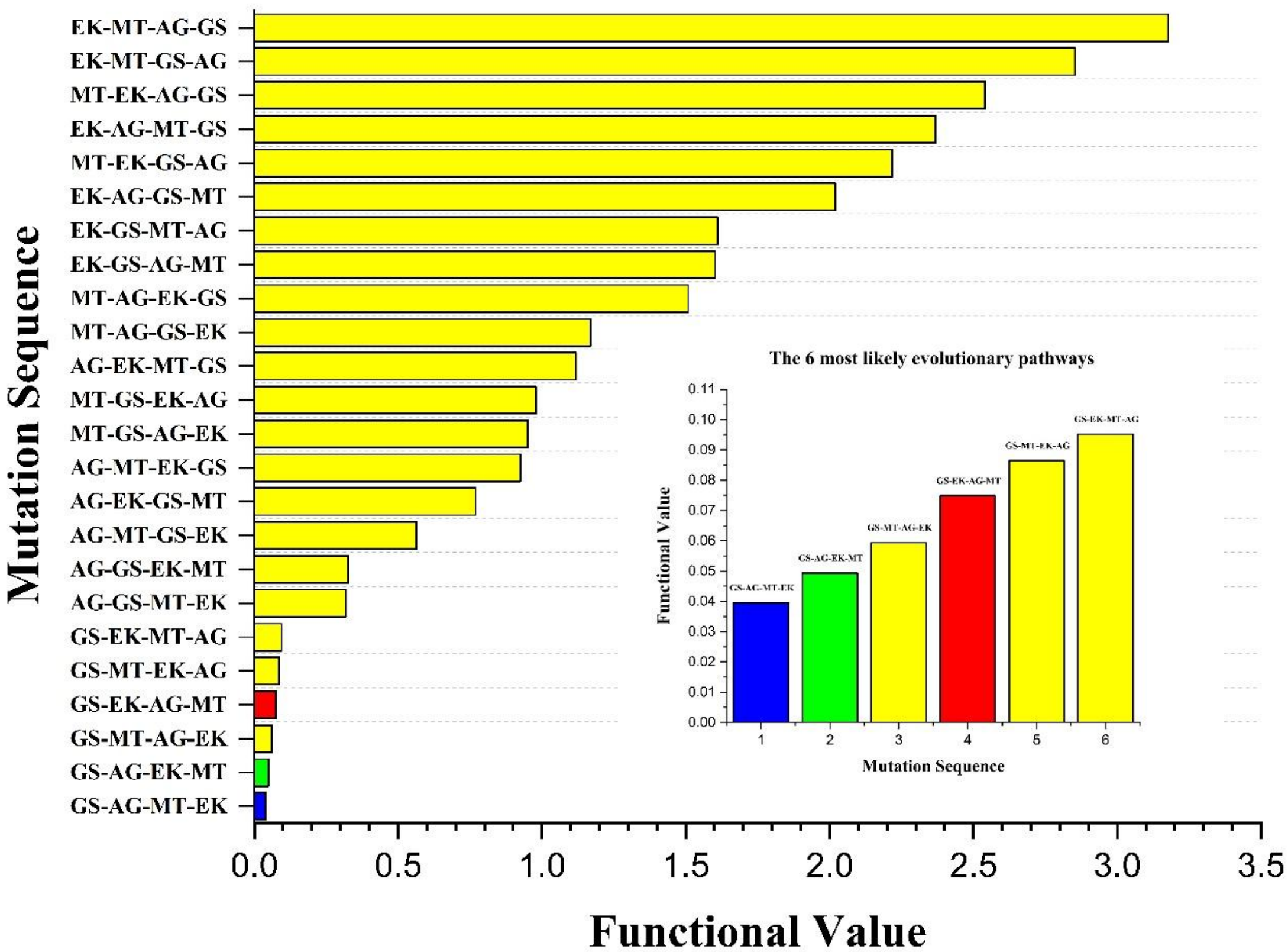

**Figure 1.** This figure illustrates the computed values of the functional, as defined in Eq. (2), for each of the 24 possible pathways resulting from four missense mutations: A42G, E104K, M182T, and G238S. For simplicity, we have omitted the specific positions of these mutations in the β-lactamase sequence from the *y*-axis. The red-filled bar indicates the most probable evolutionary pathway, as determined by Weinreich *et al*. (2006). The green-filled bar represents the second most probable pathway, based on both the findings of these authors and our calculations (see Table 1). The blue-filled bar denotes the most probable evolutionary pathway (see Table 1). In the inset, we highlight the six most probable evolutionary pathways to emphasize the differences among them, which mainly reflect the effects of epistasis.

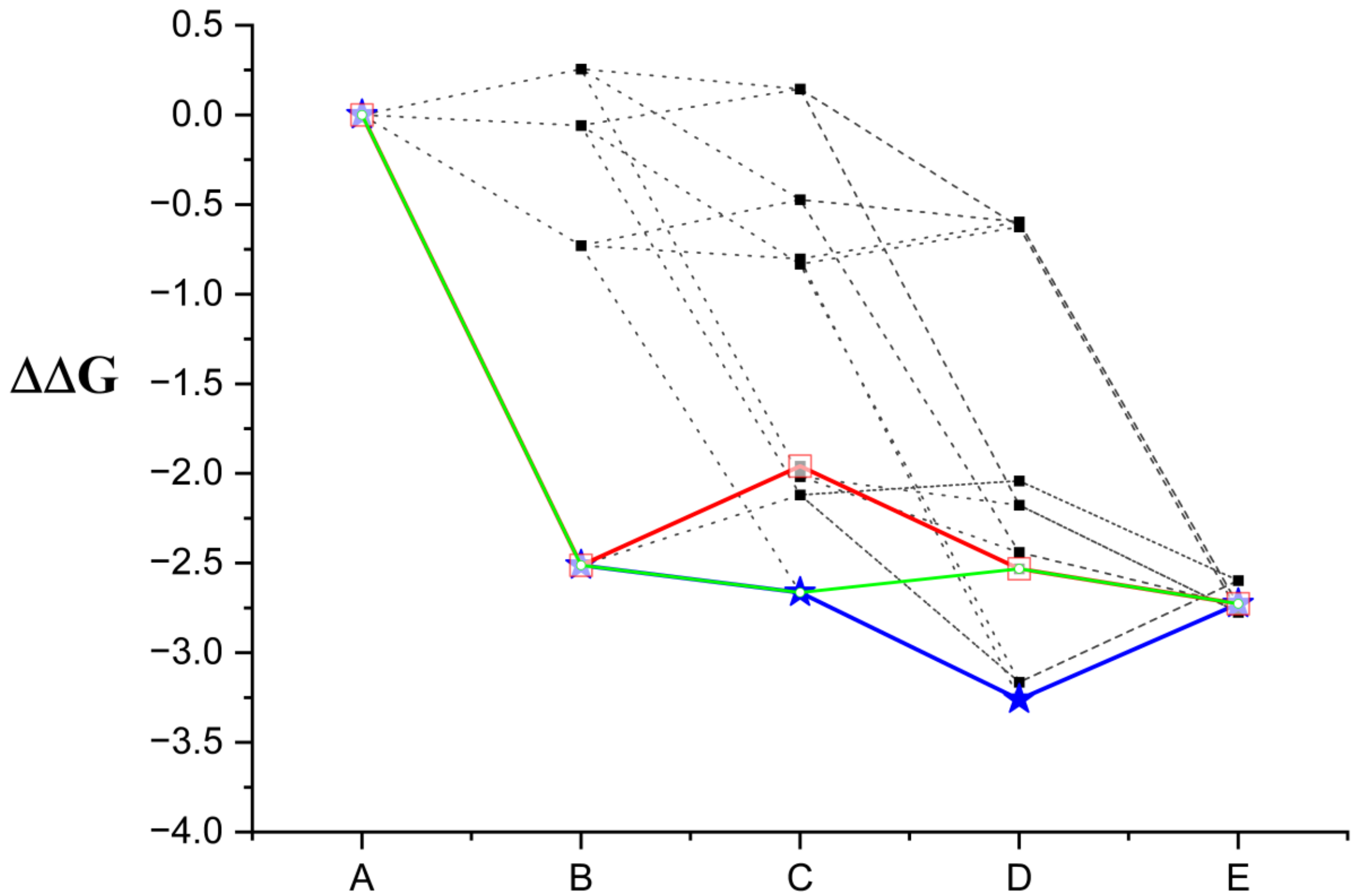

**Figure 2**. This graph displays the average free-energy change (as shown in Table 1) for each of the 24 evolutionary pathways corresponding to the four mutations. These pathways are illustrated as black dashed lines with filled black squares. The most probable pathway and, hence, the most efficient trajectory—according to Eq. (2)—is highlighted with a blue line and filled blue stars, corresponding to the following mutational order: G238S-A42G-M182T-E104K (see Table 1). The most probable pathway, as reported by Weinreich *et al*. (2006), is G238S-E104K-A42G-M182T (see Table 1), which is highlighted in red with open red squares. Additionally, the second most probable pathway—G238S-A42G-E104K-M182T (see Table 1)—according to either Weinreich *et al*. (2006) or our calculations using Eq. (2), is displayed in green lines with green circles that have white fills. In the *x*-axis, the letter **A** denotes the initial starting point for every possible pathway, specifically representing the wild-type conformation, which is defined as the one for which ΔΔG ≡ 0.0. The other letters indicate the number of mutations, which is four.